# Topological photonic states and directional emission of the light exiting from the photonic topological structure composed of two dimensional honeycomb photonic crystals with different point group symmetries

J. Hajivandi and H. Kurt

Department of Electrical and Electronics Engineering

TOBB University of Economics and Technology, Ankara 06560, Turkey

**Abstract**

In this work, we investigate a two dimensional honeycomb photonic crystal (2D HPC) with $C_6$ symmetry point group, which is known to demonstrate a double Dirac cone at $k = 0$ of the Brillion zone. Then we design two deformed PCs from the original one, by modifying the radius of the cylinders from the unit cell center in which the symmetry $C_6$ is reduced to the $C_{3v}$ group and new structure exhibits the photonic topological edge states. Consequently, the topologically protected propagation of the edge states with back scattering-immune feature is observed along the interfaces of the two deformed PCs without any defects or including cavities or bend. Furthermore, the directional surface modes exit from the photonic topological insulators (PTIs) including various defects, is investigated. As well as, we explore the propagation and coupling of light through the coupled photonic topological insulators (CPTIs).

**Introduction**

The fascinating innovation of the unique behavior of the electrons in the solid state physics, identified as the quantum spin Hall effect (QSHE) in which the topology is the main impression [1–4], provide suitable platform to train the equivalent manners in the photonic structures, like periodic PCs [5–7].

There are two approaches for stimulating the topological behavior like back-scattering immune transmission against the disorders, defects and even cavities in the PCs. The first one is breaking the time-reversal (TR) symmetry has been applied in the magnetic excitation [8–10], the dynamic modulation [11–13], and the coupled helical mechanism [14–15]. Moreover, it can be done without breaking the TR symmetry as the second approach applied in other medium such as the bi-anisotropic meta-crystals [16–23].

Dirac cones with specific features are appeared in the band structures of the two dimensional electronic, phononic and the photonic crystals are known from the first time when the effective research on the graphene structures has been done [24-36].
Several incredible wave transference properties related to the edge sates and transmission like one-way transmission and wave front shaping near the Dirac cone of the dispersion diagram of the 2 dimensional electronic [24, 37-38] and photonic medium are revealed [6, 39-44].
According to a recent study reported by L. H. Wu and X. Hu in 2015, honeycomb PCs which preserve the $C_6$ point group symmetry are one of the best lattices providing a platform for studying the robust one-way, back-scattering immune light transport, which is one of the main properties of the photonic topological insulators, without breaking the time reversal symmetry [45, 46].

In another study published in 2016, S. Barik et. al. studied the honeycomb lattice consisting of the triangular holes based on the GaAs dielectric substrate. So they perturbed the honeycomb structure by first increasing the distance between the whole holes and the center of the unit cell to obtain the expanded structure and then decreasing the radii to create a shrunken one. It was found that three dimensional honeycomb lattice consists of the triangular holes embedded on the dielectric slab with a distinct height and illuminated with the TM polarization, supports two dimensional confined helical topological edge states without more loss due to the total internal reflection of light from the slab. They also applied the circular holes in place of the triangular ones which neither band gap, nor any edge states were seen in the dispersion diagram [47].

S. Barik et. al. experimentally continued their studies. To design the topological photonic insulator, they used again previous perturbed honeycomb PCs composed of triangular holes on the GaAs substrate. By studying the dispersion diagram of this topological photonic insulator, they observed two nondegenerated helical edge states with opposite circular polarization and Poynting vectors at the bandgap region [48].

While we apply a two dimensional honeycomb PC composed of six circular rods in the air background preserve the $C_6$ point group symmetry which reveals spin quantum Hall physics associated with the two pseudospin states. We perform our perturbations by increasing and decreasing the distance between the rods and the center of the unit cell in the form of one in between which is different from the expanding or compressing the unit cell wholly. Two perturbed PCs do not preserve the $C_6$ symmetry but reduce it to $C_{3v}$ besides they show different topological behaviours. Designing a topological structure through these two materials with opposite topological behaviors, we study the robust topological edge states propagation around sharp bends, through cavities and disorders. So, the helical edge states are observed in the topological bandgap for the TM polarization in comparison with the Barik et al. structures for circular holes. As recognized early they used triangular rods instead of circular ones to detect the helical edge states.

So as we proposed, using honeycomb PC composed of circular rods is more accessible and useful in place of triangular ones from the practical point of view to study the topological features such as protected helical edge transport in the contours including bending, disorders or cavities.

As well as, recently PCs have been implemented to excite the directional emission of the surface wave at the termination of the PCs which is coupled to the air modes. We explore the directional surface mode emission of the edge modes at the end face of the PTIs [49].

Besides, PCs are the intriguing candidates for the planar and coupled cavity waveguides for producing enhanced light propagation through these types of waveguides. So, the Electro-optical switching can be regarded as the important factors for employing in the photonic integrated circuits. In this paper, we expand this kind of switching of light through the coupled photonic topological insulators (CPTIs) [50].

**Photonic Topological Insulator (PTI)**

Here, we introduce a 2D HPC where each unit cell consists of six metamaterial rods on a deeply subwavelength scale, radius $r$ with permittivity $\varepsilon = 12$ in air ($n_{rod} > n_{air}$) possessing $C_6$ symmetry, and embedded at the distance $R$ from the unit cell center with two lattice vectors $\vec{a}_1 = (a, 0)$ and $\vec{a}_2 = (\frac{a}{2}, \frac{a\sqrt{3}}{2})$ where $a$ is the lattice constant. In Figure 1(a), the graphic representation of the photonic lattice is shown.

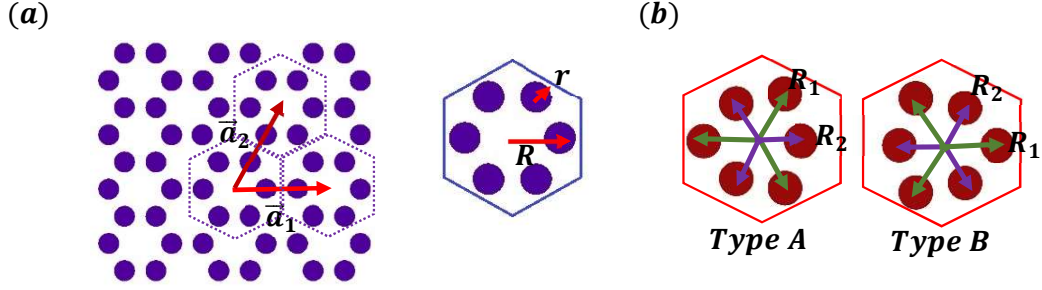

Fig. 1. (a) The graphic geometry of the 2D HPC with a hexagonal unit cell composed of 6 rods with radius $r$ embedded in the distance $R$ from the center, $\vec{a}_1$ and $\vec{a}_2$ are the lattice vectors, (b) The modified PCs types $A$ and $B$. In both types, the green vectors indicate the distance $R_1 = 0.35a$ and the purple vectors indicate the distance $R_2 = 0.28a$.

Figure 2(a) indicates the dispersion diagram of the 2D HPC in which $r = 0.1111a$ and $R = 0.3333a$ for the selected transverse magnetic (TM) mode where $(E_z, H_x, H_y) \neq 0$ and $(H_z, E_x, E_y) = 0$ [45]. So the double Dirac cone which will appear at the central point of the Brillion zone, $\varGamma$, indicates the four-fold degeneracies between four dipolar and quadrupolar states, $p_x/p_y$ and $d_{x^2-y^2}/d_{xy}$ due to the $C_6$ symmetry. In fact, one may employ these representations to form pseudo-time-reversal symmetry. We perform the numerical calculations of the photonic band structures via the package *MIT Photonics Bands* (MPB) [51].

Here we note that the behavior of honeycomb PC in which the Dirac cone is revealed at the central point, $\varGamma$ is similar to the met crystal with zero effective refractive index $n_{eff}(\omega) = 0$ [36].

Breaking the symmetry of the PC leads to the QSHE, supplemented with the topologically edge states seen between regions with different bulk topological phases, which is the most fascinating properties of the photonic topological insulators. By modifying the distance $R$, the coupling between adjacent metamolecules will be changed, ultimately results in reducing the point group symmetry $C_6$ to $C_{3v}$, lifting the Dirac point, so appearing the two-fold degenerated bands. Figure 2(b) represents the common band diagram of both modified PCs types $A$ and $B$ with different topological behaviors.

The electric field profiles of the two-folded degenerate bands at the $\varGamma$ point of types $A$, $B$ and 2DHPC, shown in the Figure 2(c) where they are symmetric and antisymmetric modes. Moreover, the field profiles of PCs types, $A$ and $B$ convert to each other through the blue vectors. This behavior which is associated with band inversion besides closing-opening the Dirac points provide an impression to study back scattering-immune unidirectional propagation through the topological phase transitions. For performing this purpose, we design a left/right structure with a supercell made of two PCs kinds $A$ & $B$. Figure 3. indicates the appeared helical edge states in the band gap region which is same as one for the unit cell of types $A$ and $B$ seen in the Figure 2(b).

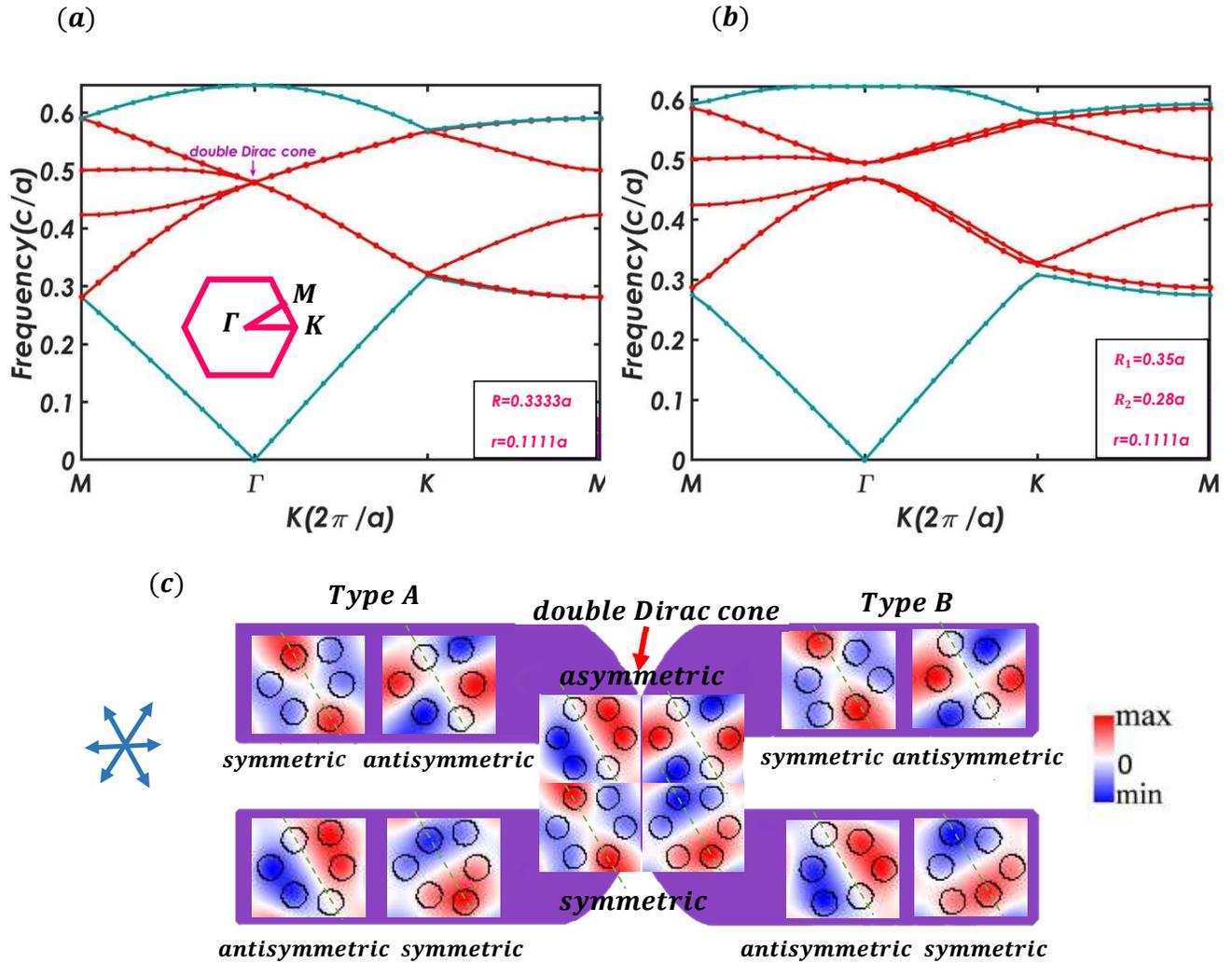

Fig. 2. (a) The dispersion diagram of a 2D honeycomb meta PC where rods are at $\varepsilon = 12$, $R = 0.3333a$ and $r = 0.1111a$ for the TM polarization, the inset indicates the Brillion zone of a 2D honeycomb PC, (b) the band diagram of the modified types *A* and *B*, (c) the profiles of the z component of the electric fields for the degenerated bands in (a) and (b) for types *A* & *B*. The field profiles of types *A* and *B* are converted to each other along the blue vectors.

As seen in the Figures 3(e) and (f), two edge states 1 and 2, named spin up and down respectively, have same electric field profiles, but opposite Poynting vectors. Another design composed of PCs, types *A* and *B* is indicated in the Figure 3(b). As seen in the Figure 3(d), just the bulk states are appeared. In other words, there is a complete band gap.

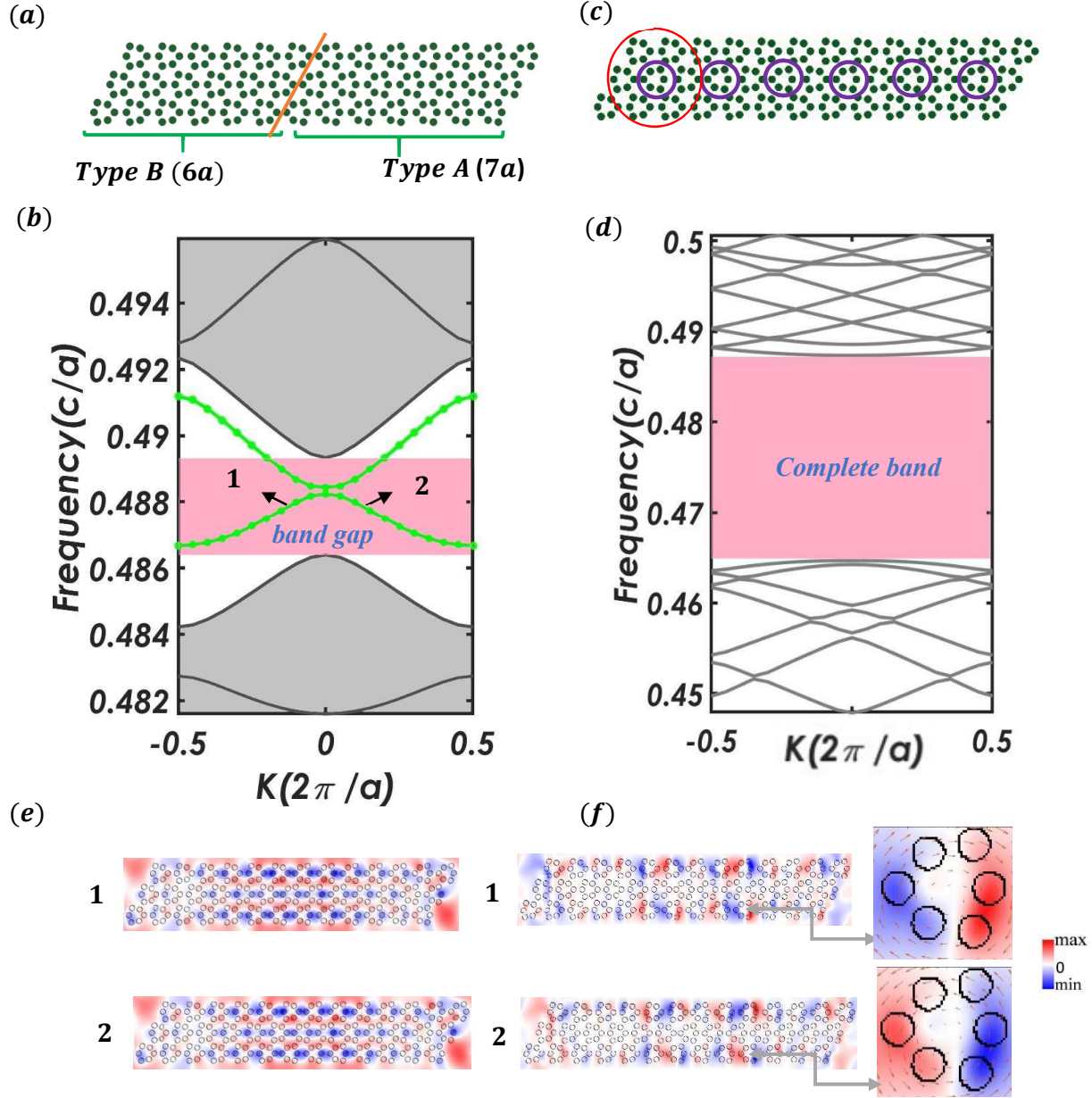

Fig. 3. (a) The graphic geometry of the supercell made of two PCs types *A* and *B* with oblique boundary (orange line) and (b) its band diagram with two edge states (green bands), (c) other geometry of supercell without continuous boundary (purpule circles in the big red circle) and (d) its band diagram in the absence of the any edge states, (e) two electric field profiles of helical states, **1** and **2**, (f) the Poynting vector profiles of edge states, **1** and **2**.

To verify the unidirectional propagation of light and the negligible attenuation around the sharp corners (bends) and cavities, we apply the *Lumerical FDTD solution* [52] to design an up-down photonic topological insulator (PTI), composed of metamaterial honeycomb PCs, types *A* and *B* and we use the PML boundary conditions around the structure, Figure 4. Two magnetic dipoles at the edge state frequencies are embedded in the interface of two PCs, types *A* and *B*, to produce electric field $E_z$, (yellow star). They are vertical and $90^0$ phase differences are between them ( $H_x + iH_y$ ).

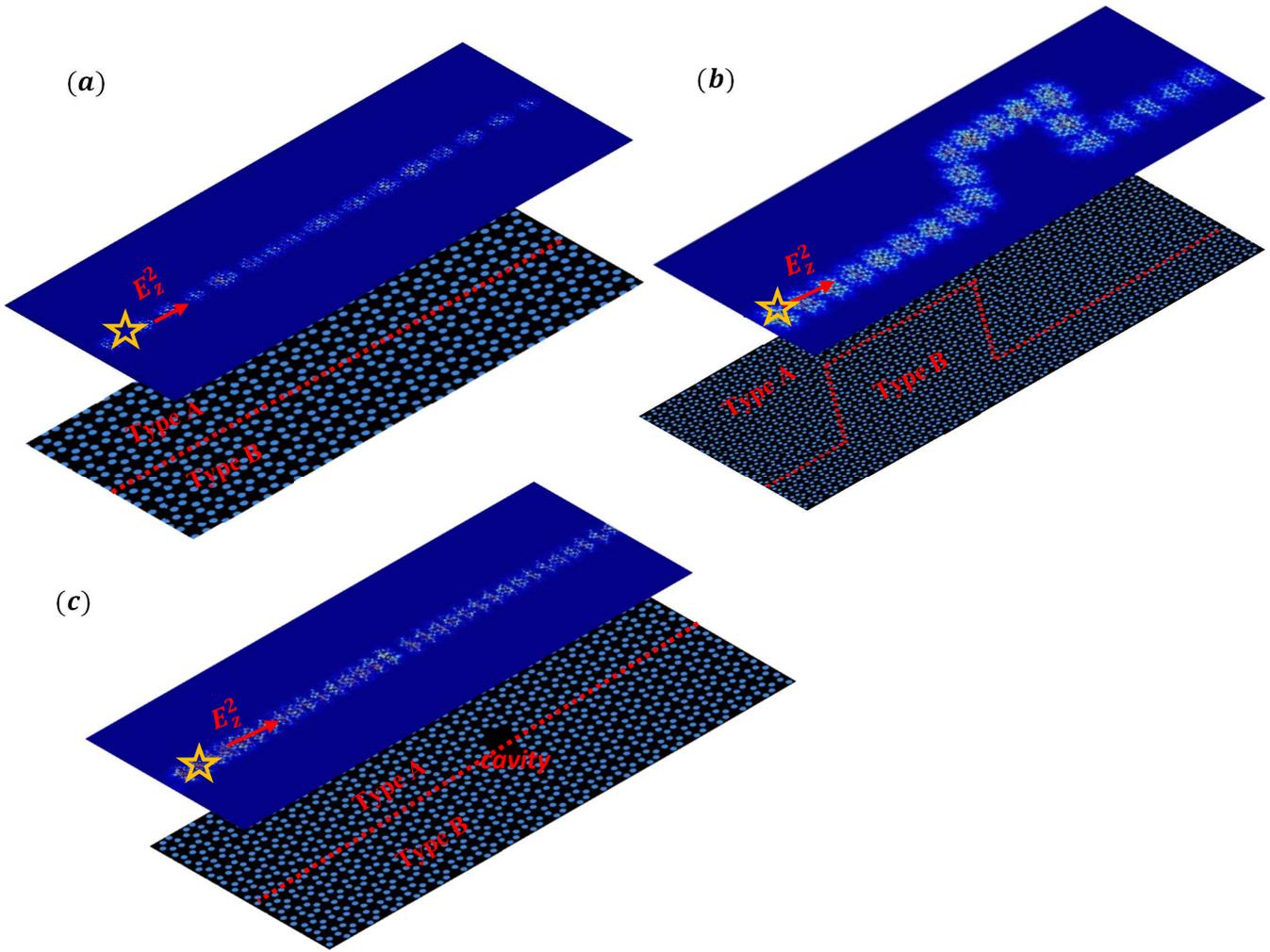

Fig. 4. (a) Propagation of the edge sates in one-way at the interface of two types of PCs, *A* and *B* without any defects and propagation with backscattering immunity at (b) arbitrary bend defect, (c) an arbitrary-shaped cavity made by removing the rods of two unit cells, at the common boundary, the exact locations of two orthogonal magnetic dipoles is shown by the star.

**Directional Surface Mode Emission Using Photonic Topological Insulators**

We explore the directional emission of the light exiting the photonic topological structure. Indeed, the beam divergence takes place when the illuminated beam of the wavelength $\lambda$, leaves the structure of the subwavelength geometry Removing this unfavorable effect, so reaching the directional surface mode emission have been considerable recently. For example, applying various defects at a photonic crystal may lead to achieving the directional edge states.

Here, we investigate highly directional topological edge mode exiting the PTI applying various kinds of defects through our structure. In addition, we explore the enhanced transmission of the topological directional mode in comparison with the first topological diverged mode at the distances far from the exit of the PTI.

At first, we study the topological edge mode with the frequency, $f = 190\ THz$, exits from our previous up/down design.

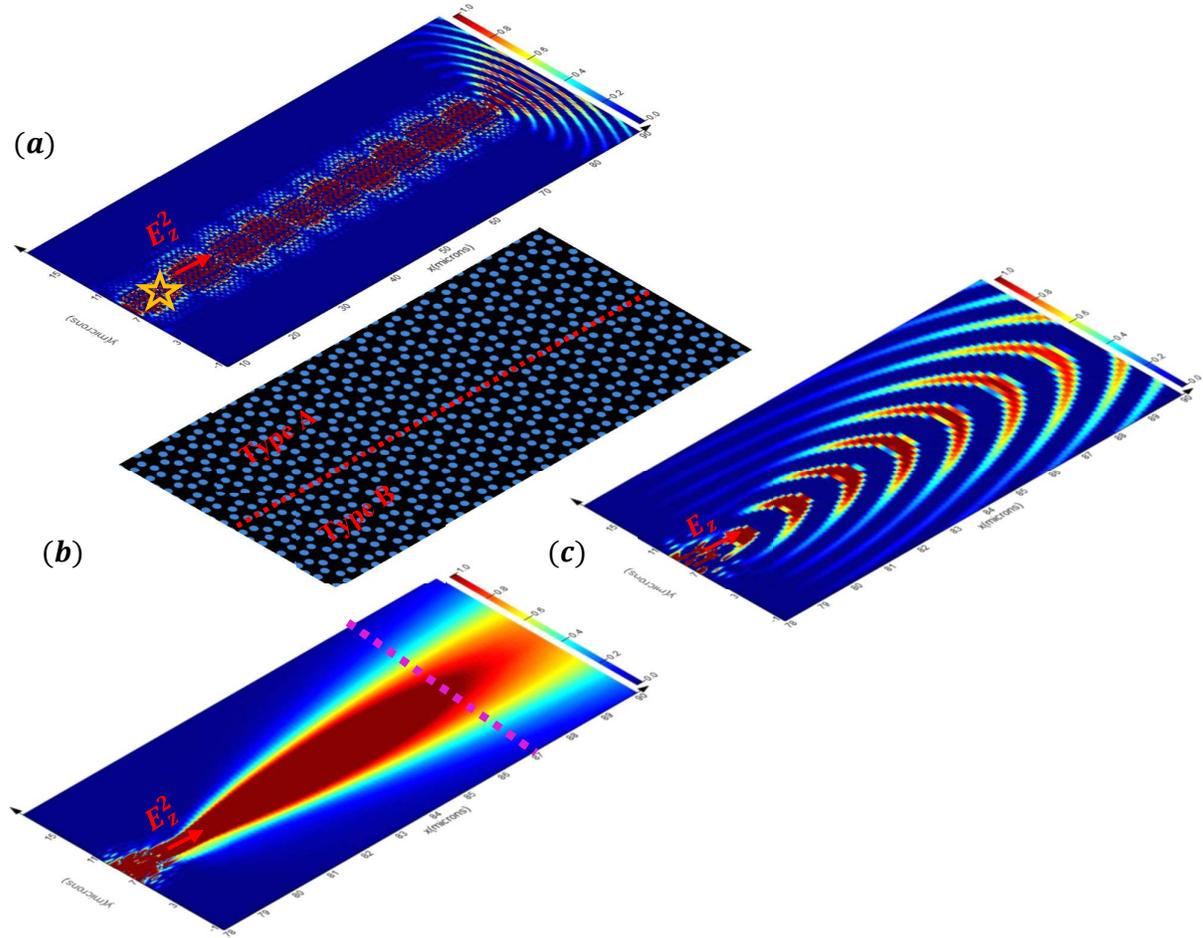

Fig. 5. (a) Propagation of the edge sate inside and outside the PTI, (b) the intensity profile of the edge state, $E_z^2$, in front of the PTI output, (c) profile of real part of the electric field, $E_z$, in front of the PTI output.

As seen in the Figure 5. (a), the one-way edge state propagating from the source, experiences a high divergence leaving the topological structure, so entering the free space. Figures 5. (b)-(c), show the intensity, $|E_z|^2$ and $real\ (E_z)$ at the free space region in front of the PTI respectively.

We design various cavity defects at termination of the PTI in order to manufacture the directional edge states out of the topological geometry.

Figure 6. (a), shows the directional edge mode leaving the PTI manufactured by creating a cavity at the region of the PC type A. Indeed, this cavity has been made by removing one unit cell of PC, type A. We found that the output light of the recent structure becomes narrower than the light of the main structure in Figure 5. (a). Besides, the light intensity of the defected

structure becomes more powerful than the light intensity of the main structure at the distances greater than 87 micrometers, indicated with pink dashed line).

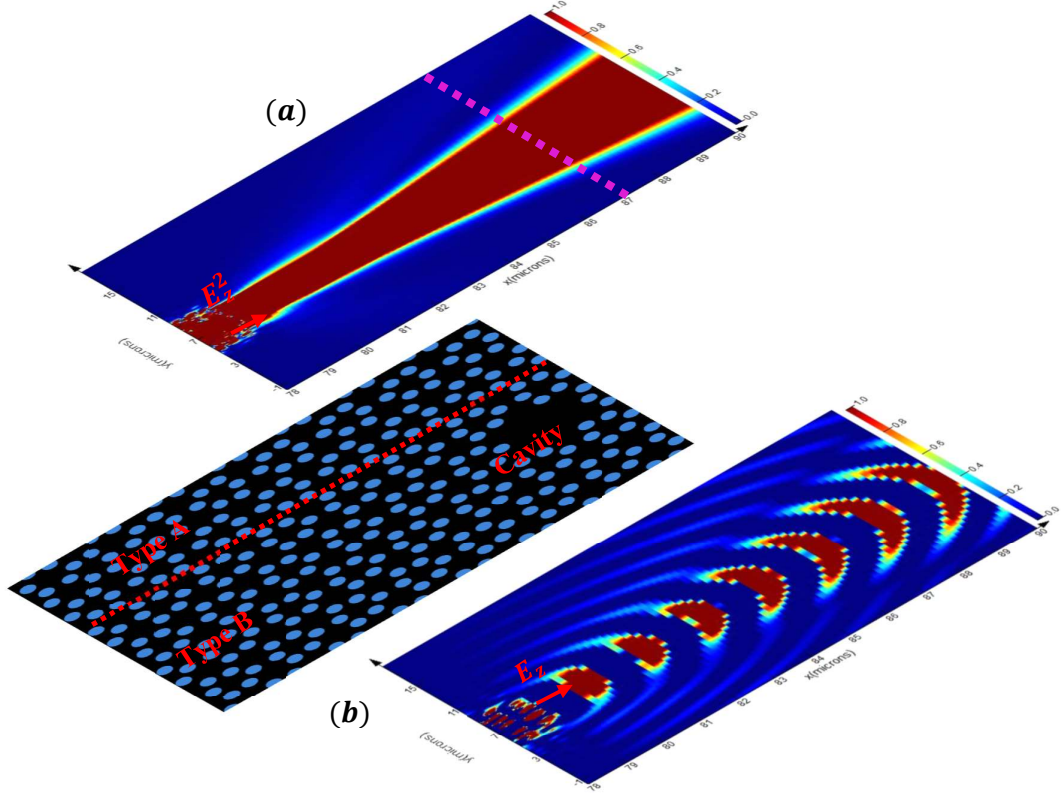

Fig. 6. Propagation of the edge sate outside the defected PTI including one cavity, (a) the intensity profile of the edge state, $E_z^2$, in front of the defected PTI output, (b) profile of real part of the electric field, $E_z$, in front of the defected PTI output.

Moreover, the light divergence may be decreased by creating other cavity in the PC, type B. For studying the behavior of the topological edge state, $f = 190\ THz$, in front of the output, we insert various line detectors at different points in front of the main and last defected PCs exit. As seen in the Figure (), although the transmission of light of the defected PTI is about 65% (0.75/1.15) of the main PTI at the exit, $x = 80\mu m$, it becomes higher than the main PTI at $x \geq 100\mu m$.

We also study the transmission of other modes (not just the edge mode), in different positions in front of the main PTI ($T_1$, $T_2$, $T_3$, $T_4$, $T_5$) and defected PTI ($T_6$, $T_7$, $T_8$, $T_9$, $T_{10}$), see in the Figure 8 (b). As seen in the Figure 8 (c). for he edge mode, $f \cong 190\ THz$, the transmissions of the light in front the defected PTI ($T_6$, $T_7$, $T_8$, $T_9$, $T_{10}$), become higher than the main PTI ($T_1$, $T_2$, $T_3$, $T_4$, $T_5$) which is in the overlap with the Figure 8 (b).

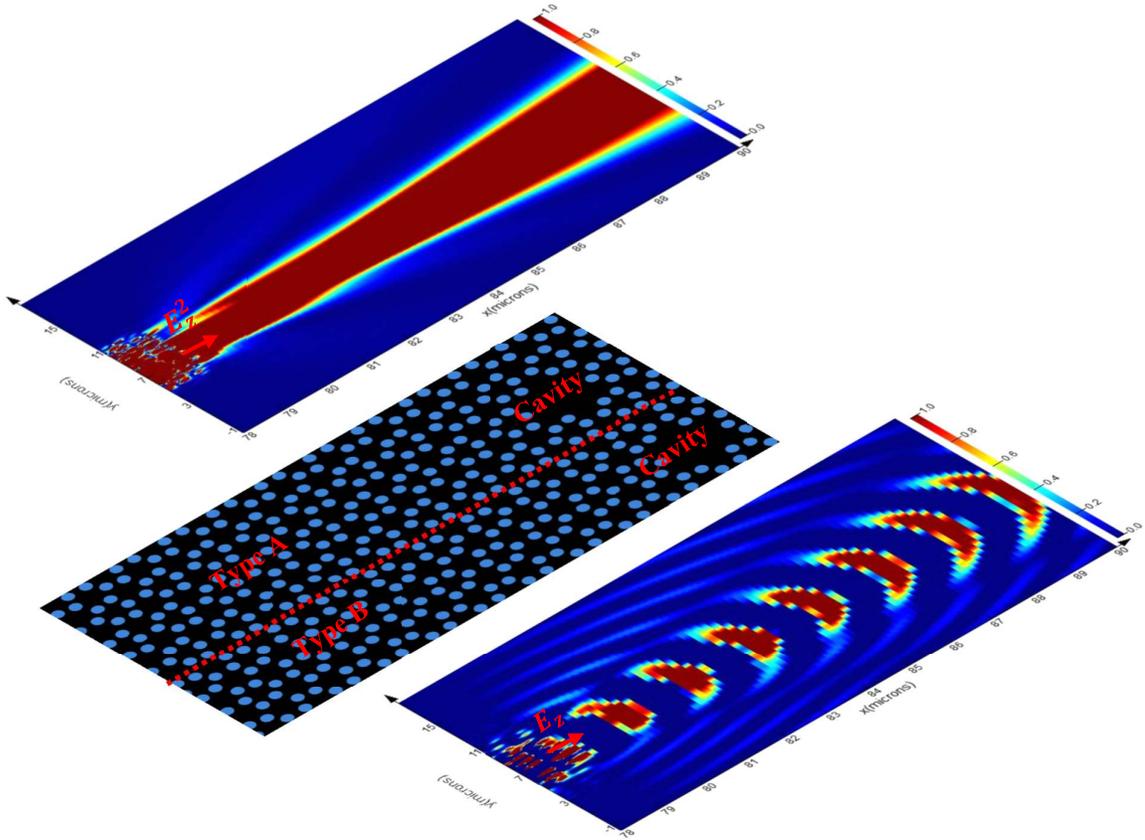

Fig. 7. Propagation of the edge sate outside the defected PTI including two cavities, (a) the intensity profile of the edge state, $E_z^2$, in front of the defected PTI output, (b) profile of real part of the electric field, $E_z$, in front of the defected PTI output.

## Coupled Photonic Topological Insulators (CPTIs)

Inserting two two-dimensional PC waveguides in closeness of each other, provides a fascinating approach for in-plane propagating and coupling of the light through this kind of directional coupler electro-optical device. The coupled wave can turn back into the first waveguide which lead to exchanging the power alternatively.

Here, we prospect a light switching performance in coupled photonic topological interfaces. Indeed, the switch is carried out between two interfaces in PTIs composed of PCs types A and B. As seen in the Figure 9 (b) (center), light propagates from the input in the down PTI, strongly couples to the upper PTI through the interfaces and exit from the outputs. The transported light to the outputs 1 and 3 is so weak in comparison with the output 2, Figure 9 (b) (right and left sides).

Besides, the light transportation is studies through the different points of the CPTIs, in lines 1-4. Figures 9(c)-(f) clearly, indicate that through the line 1, the light propagates from $x = 15 \mu m$. So the transported light to the right side is a few in comparison with the source intensity. Decreasing of the intensity of light begins from the middle of line 2. Meanwhile, increasing the intensity starts from the middle of the line 3. The observed pattern of intensity, for the line 4, is opposite with whatever seen through the line 1. These patterns obviously, show the light switching performance through the CPTIs. In addition, in Figure 9 (g), the maximum intensity of the point 2 near the edge mode frequency, is shown in comparison with the points 1, 3

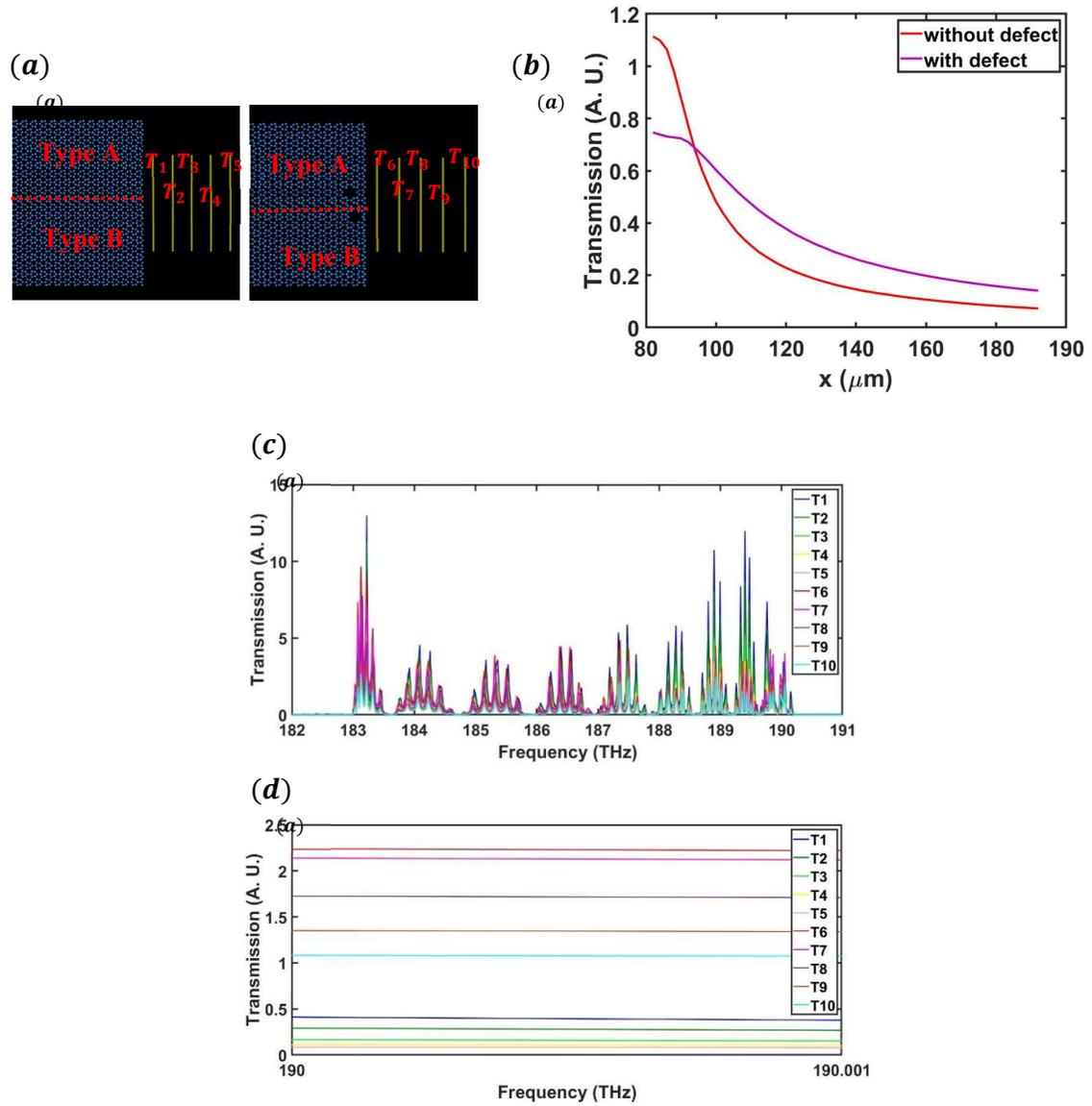

Fig. 8. (a) Geometry of the various detectors in front of the PTI (left) and defected PTI including two cavities (right), (b) transmission power in front of the PTI and defected PTI, (c) transmission power detected by the monitors in (a) for various frequencies, (d) transmission power detected by the monitors in (a) for the frequencies near the edge mode.

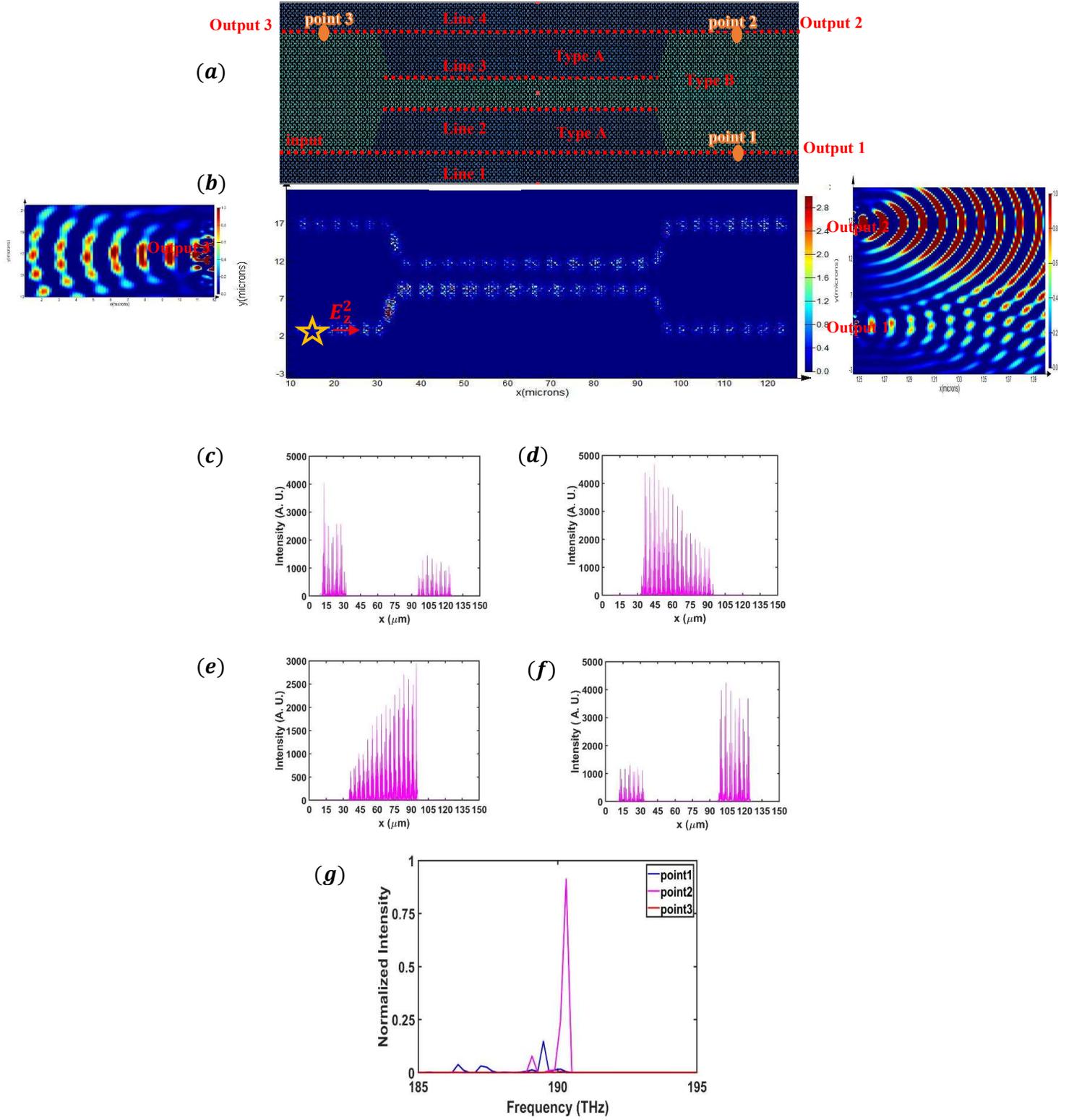

Fig. 9. (a) Geometry of the coupled photonic topological insulators (CPTIs) composed of PCs, types A and B, (b) profile of intensity of the electric field, $E_z^2$, for the edge mode, through the CPTIs (center) and outside the CPTIs. In output 1.2.3 (right and left) , (c)-(f) intensity of the electric field, $E_z^2$, for the edge mode, versus x through the lines 1-4, (g) intensity of the electric field, $E_z^2$, versus frequency at the points 1,2,3.

## Conclusion

In this paper, we study a 2D HPC in which the Dirac cone is observed at the central point of the Brillion zone, This PC acts as a zero refractive index metamaterial. Besides by modifying the radius of the cylinders from the center of the unit cell, two types of modified PC are proposed which can be used for designing the photonic topological insulators. The photonic edge states are secured topologically and robust against the disorders, bends and even cavities. Besides, we investigate highly directional topological edge mode exiting from the PTI including various kinds of defects. In addition, we explore the enhanced transmission of the topological directional mode in comparison with the first topological diverged mode at the distances far from the exit of the PTI

## Acknowledgment

H. K. acknowledges partial support of the Turkish Academy of Sciences.

## References

[1] M. Konig, S. Wiedmann, C. Brune, A. Roth, H. Buhmann, L.W. Molenkamp, X. L. Qi and S. C. Zhang, Science **318**, 766 (2007).

[2] B. A. Bernevig, T. L. Hughes and S. C. Zhang, Science **314**, 1757 (2006).

[3] M. Z. Hasan and C. L. Kane, Rev. Mod. Phys. **82**, 3045 (2010).

[4] D. Hsieh, D. Qian, L. Wray, Y. Xia, Y. S. Hor, R. J. Cava, and M. Z. Hasan, Nature (London) **452**, 970 (2008).

[5] L. Lu, J. D. Joannopoulos and M. Soljaclc, Nat. Photonics **8**, 821 (2014).

[6] F. D. M. Haldane and S. Raghu, Phys. Rev. Lett. **100**, 013904 (2008).

[7] B. Z. Xia, T. T. Liu, G. L. Huang, H. Q. Dai, J. R. Jiao, X. G. Zang, D. J. Yu, S. J. Zheng and J. Liu, J. Appl. Phys. **122**, 065103 (2017).

[8] Z. Wang, Y. D. Chong, J. D. Joannopoulos and M. Soljacic, Phys. Rev. Lett. **100**, 013905 (2008).

[9] Z. Wang, Y. Chong, J. D. Joannopoulos and M. Soljacic, Nature (London) **461**, 772 (2009).

[10] C. He, X. L. Chen, M. H. Lu, X. F. Li, W. W. Wan, X. S. Qian, R. C. Yin and Y. F. Chen, Appl. Phys. Lett. **96**, 111111 (2010).

[11] F. Gao, Z. Gao, X. Shi, Z. Yang, X. Lin, H. Xu, J. D. Joannopoulos, M. Soljacic, H. Chen, L. Lu, Y. Chong, and B. Zhang, Nat. Commun. **7**, 11619 (2016).

[12] W. J. Chen, S. J. Jiang, X. D. Chen, B. Zhu, L. Zhou, J. W. Dong, and C.T. Chan, Nat. Commun. **5**, 6782 (2014).

[13] C. He, X. C. Sun, X. P. Liu, M. H. Lu, Y. Chen, L. Feng, and Y. F. Chen, Proc. Natl. Acad. Sci. USA **113**, 4924–4928 (2016).

[14] F. F. Li, H. X. Wang, Z. Xiong, Q. Lou, P. Chen, R.X. Wu, Y. Poo, J. H. Jiang & S. John, Nature Commu. **9**, 2462 (2018).

[15] Y. E. Kraus, Y. Lahini, Z. Ringel, M. Verbin & O. Zilberberg, Phys. Rev. Lett. **109**, 106402 (2012).

[16] Y. E. Kraus, Z. Ringel & O. Zilberberg, Phys. Rev. Lett. **111**, 226401 (2013).

[17] O. Zilberberg, S. Huang, J. Guglielmon, M. Wang, K. P. Chen, Y. E. Kraus and M. C. Rechtsman, Nature **553**, 59–62 (2018).

[18] M. Lohse, C. Schweizer, H. M. Price, O. Zilberberg & I. Bloch, Nature **553**, 55–58 (2018).

[19] P. S. Jean, V. Goblot, E. Galopin, A. Lemaître, T. Ozawa, L. Le Gratiet, I. Sagnes, J. Bloch and A. Amo, Nat. Photon. **11**, 651–656 (2017).

[20] A. P. Slobozhanyuk, A. N. Poddubny, A. E. Miroshnichenko, P. A. Belov & Y. S. Kivshar, Phys. Rev. Lett. **114**, 123901 (2015).

[21] X. L. Qi & S. C. Zhang, Rev. Mod. Phys. **83**, 1057–1110 (2011).

[22] Z. L. Xiang, S. Ashhab, J. Q. You & F. Nori, Rev. Mod. Phys. **85**, 623–653 (2013).

[23] Y. Ran, Y. Zhang & A. Vishwanath. Nat. Phys. **5**, 298–303 (2009).

[24] K. S. Novoselov, A. K. Geim, S. V. Morozov, D. Jiang, M. I. Katsnelson, I. V. Grigorieva, S. V. Dubonos and A. A. Firsov, Nature **438**, 197 (2005).

[25] D. Torrent, D. Mayou and J. Sanchez-Dehesa, Phys. Rev. B **87**, 115143 (2013).

[26] X. Zhang, Phys. Rev. Lett. **100**, 113903 (2007).

[27] X. Zhang and Z. Liu, Phys. Rev. Lett. **101**, 264303 (2008).

[28] D. Malko, C. Neiss, F. Vines and A. Gorling, Phys. Rev. Lett. **108**, 086804 (2012).

[29] D. Torrent and J. S anchez-Dehesa, Phys. Rev. Lett. **108**, 174301 (2012).

[30] J. Lu, C. Qiu, S. Xu, Y. Ye, M. Ke and Z. Liu, Phys. Rev. B **89**, 134302 (2014).

[31] J. Lu, C. Qiu, L. Ye, X. Fan, M. Ke, F. Zhang and Z. Liu, Nat. Phys. **13**, 369–374 (2017).

[32] Y. Li, Y. Wu and J. Mei, Appl. Phys. Lett. **105**, 014107 (2014).

[33] Y. Li and J. Mei, Opt. Express **23**, 12089 (2015).

[34] T. Ochiai and M. Onoda, Phys. Rev. B **80**, 155103 (2009).

[35] S. H. Kim, S. Kim and C. S. Kee, Phys. Rev. B **94**, 085118 (2016).

[36] X. Huang, Y. Lai, Z. H. Hang, H. Zheng and C. T. Chan, Nat. Mater. **10**, 582 (2011).

[37] Y. Zhang, Y. W. Tan, H. L. Stormer and P. Kim, Nature **438**, 201-204 (2005).

[38] A. H. C. Neto, F. Guinea, N. M. R. Peres, K. S. Novoselov and A. K. Geim, Rev. Mod. Phys. **81**, 109-162 (2009).

[39] L. G. Wang, Z. G. Wang, J. X. Zhang and S. Y. Zhu, Opt. Lett. **34**, 1510-1512 (2009).

[40] R. A. Sepkhanov, Y. B. Bazaliy and C. W. J. Beenakker, Phys. Rev. A **75**, 063813 (2007).

[41] X. Zhang, Phys. Rev. Lett. **100**, 113903 (2008).

[42] M. Diem, T. Koschny and C. M. Soukoulis, Physica B **405**, 2990-2995 (2010).

[43] M. Plihal and A. A. Maradudin, Phys. Rev. B **44**, 8565-8571 (1991).

[44] S. Raghu and F. D. M. Haldane, Phys. Rev. A **78**, 033834 (2008).

[45] L.H. Wu and X. Hu, Phys. Rev. Lett. **114**, 223901 (2015).

[46] J. Hajivandi and H. Kurt, arXiv:2002.00588 (2020).

[47] S. Barik, H. Miyake, W. DeGottardi, E. Waks and M. Hafezi, New J. Phys. **18,** 113013 (2016).

[48] S. Barik et.al., Science **359**, 666–668 (2018).

[49] Q. Wang, Y. Cui1, C. Yan, L. Zhang and J. Zhang, J. Phys. D: Appl. Phys. **41**, 105110 (2008).

[50] A. Sharkawy, S. Shi and D. W. Prather, Opt. Exp. **10**, 1048 (2002).

[51] S.G. Johnson and J.D. Joannopoulos, Opt. Express **8**, 173 (2001).

[52] Lumerical Inc. http://www.lumerical.com/tcad-products/fdtd/